\newif\ifcol
\coltrue

\newif\ifbw
\bwfalse

\newif\ifdraft
\draftfalse

\newcommand{\pagesize}{a4paper}

\ifdraft
  \colfalse
\fi

\newlength{\mapwidth}
\ifdraft
\fi

\newlength{\coeffwidth}
\ifdraft
  \documentclass[11pt,draftcls,\pagesize,onecolumn,twoside]{IEEEtran}
\else
  \documentclass[10pt,journal,\pagesize,twoside]{IEEEtran}
\fi

\usepackage{url}

\usepackage{ifpdf}

\ifpdf
  \usepackage[pdftex]{hyperref}
\else
  \usepackage[ps2pdf,colorlinks=true]{hyperref}
\fi

\usepackage{cite}      

\ifpdf
  \usepackage[pdftex]{graphicx}
\else
  \usepackage[dvips]{graphicx}
\fi

\usepackage{subfigure} 

\usepackage{amssymb}    

\newcommand{\el}{\ensuremath{\ell}}
\newcommand{\m}{\ensuremath{m}}
\newcommand{\n}{\ensuremath{{n}}}
\newcommand{\elm}{\ensuremath{{\el \m}}}

\newcommand{\elmax}{\ensuremath{{\el_{\rm max}}}}
\newcommand{\dmatbig}{\ensuremath{D}}
\newcommand{\dmatsmall}{\ensuremath{d}}
\newcommand{\admissc}{\ensuremath{C}}
\newcommand{\admissC}{\ensuremath{\admissc_\wavm^\el}}

\newcommand{\shc}[3]{\ensuremath{{#1}_{#2#3}}}
\newcommand{\shcc}[3]{\ensuremath{{#1}_{#2#3}^\cconj}}

\newcommand{\shfarg}[3]{\ensuremath{Y_{#1#2}({#3})}}
\newcommand{\shfargc}[3]{\ensuremath{Y_{#1#2}^\cconj({#3})}}
\newcommand{\shf}[2]{\ensuremath{Y_{#1#2}}}
\newcommand{\leg}[2]{\ensuremath{P_{{#1}}({#2})}}
\newcommand{\aleg}[3]{\ensuremath{P_{#1}^{#2}({#3})}}

\newcommand{\Dlmnpc}{\ensuremath{ \dmatbig_{\m\n}^{\el\cconj}(\eul) }}

\newcommand{\zreal}{{\ensuremath{\rm{Re}}}}
\newcommand{\zimag}{{\ensuremath{\rm{Im}}}}

\newcommand{\cmb}{{CMB}}
\newcommand{\cmbtext}{{cosmic microwave background}}

\newcommand{\eqn}[1]
	{(#1)}

\newcommand{\fig}[1]
	{Fig.~#1}

\newcommand{\sectn}[1]
	{section~#1}

\newcommand{\eg}{\mbox{\it{e.g.}}}

\newcommand{\cswt}
	{{CSWT}}

\newcommand{\xvect}{\ensuremath{\mathbf{x}}}

\newcommand{\sky}{\ensuremath{f}}
\newcommand{\skywav}{\ensuremath{{\mathcal{W}_\wavm^\sky}}}
\newcommand{\skyLwav}{\ensuremath{{\mathcal{W}_\wavm^{L_{\wavm}\sky}}}}

\newcommand{\spcend}{\ensuremath{\:}}

\newcommand{\img}{\ensuremath{\mathit{i}}}
\newcommand{\dx}{\ensuremath{\mathrm{\,d}}}
\newcommand{\dmu}{\ensuremath{\dx \Omega}}
\newcommand{\deul}{\ensuremath{\dx \eul}}

\newcommand{\cconj}{\ensuremath{\ast}}  
\newcommand{\sa}{\ensuremath{\omega}}
\newcommand{\saa}{\ensuremath{\theta}}
\newcommand{\sab}{\ensuremath{\phi}}
\newcommand{\sas}{\ensuremath{\saa, \sab}}
\newcommand{\eul}{\ensuremath{\mathbf{\rho}}}
\newcommand{\euls}{\ensuremath{\eula, \eulb, \eulc}}
\newcommand{\eula}{\ensuremath{\alpha}}
\newcommand{\eulb}{\ensuremath{\beta}}
\newcommand{\eulc}{\ensuremath{\gamma}}

\newcommand{\realno}{\ensuremath{\mathbb{R}}}
\newcommand{\intno}{\ensuremath{\mathbb{Z}}}
\newcommand{\natno}{\ensuremath{\mathbb{N}}}
\newcommand{\sothree}{\ensuremath{{\mathrm{SO}(3)}}}
\newcommand{\sphere}{\ensuremath{{S^2}}}

\newcommand{\rot}{\ensuremath{\mathcal{R}}}
\newcommand{\scalea}{\ensuremath{R}}

\newcommand{\p}{\ensuremath{^\prime}}

\newcommand{\wav}{\ensuremath{\Psi}}
\newcommand{\wavm}{\ensuremath{\Upsilon}}
\newcommand{\wavmc}{\ensuremath{\Upsilon^\cconj}}

\newcommand{\sumlm}{\ensuremath{\sum_{\el=0}^{\infty} \sum_{\m=-\el}^\el}}
\newcommand{\sumlmn}{\ensuremath{\sum_{\el=0}^{\infty} \sum_{\m=-\el}^\el} \sum_{\n=-\el}^\el}

\newcommand{\kron}{\ensuremath{\delta}}

\newcommand{\etal}{\mbox{\it{et al.}}}

\begin{document}
\title{A directional continuous wavelet\\transform on the sphere}
%
%

%

\author{J.~D.~McEwen,~M.~P.~Hobson,~A.~N.~Lasenby
\thanks{Manuscript received 20 April, 2006}
\thanks{The authors are with the 
        Astrophysics Group, Cavendish Laboratory, Cambridge, UK.}
\thanks{E-mail: mcewen@mrao.cam.ac.uk (J.~D.~McEwen)}}
%
%
%
\markboth{IEEE Transactions on Signal Processing,~Vol.~--, No.~--,~April~2006}%
{McEwen \MakeLowercase{\textit{et al.}}: A directional continuous wavelet transform on the sphere}
%



\maketitle



\begin{abstract}
A new construction of a directional continuous wavelet analysis on the sphere is derived herein.  We adopt the harmonic scaling idea for the spherical dilation operator recently proposed by Sanz \etal\ \cite{sanz:2006} but extend the analysis to a more general directional framework.  Directional wavelets are a powerful extension that allow one to also probe oriented structure in the analysed function. 
Our spherical wavelet methodology has the advantage that all functions and operators are defined directly on the sphere.
The construction of wavelets in our framework is demonstrated with an example.
\end{abstract}


\IEEEpeerreviewmaketitle


\section{Introduction}
\label{sec:intro}

\PARstart{T}{he} 
extension of wavelet analysis to a spherical manifold is of important theoretical and practical interest.  Wavelets are a powerful signal analysis tool due to the spatial and scale localisation encoded in the analysis.  The usefulness of such an analysis has previously been demonstrated on a large range of physical applications. 
However, many of these applications are restricted to Euclidean space $\realno^n$, where the dimension of the space $n$ is often one or two.  Nevertheless, data are often defined on other manifolds, such as the 2-sphere (\sphere; hereafter we refer to the 2-sphere simply as the sphere).  
For example, applications where data are defined on the sphere are found in 
astrophysics (\eg\ \cite{bennett:1996,bennett:2003}), 
planetary science (\eg\ \cite{wieczorek:2006,wieczorek:1998,turcotte:1981}), 
geophysics (\eg\ \cite{whaler:1994,swenson:2002,simons:2006}), 
computer vision (\eg\ \cite{ramamoorthi:2004}) and 
quantum chemistry (\eg\ \cite{choi:1999,ritchie:1999}).
A wavelet transform on the sphere is required to realise the potential benefits of a wavelet analysis in these settings.
In this work we construct a new directional continuous spherical wavelet transform (\cswt).

A number of attempts have been made previously to extend wavelets to the sphere.
Discrete second generation wavelets on the sphere that are based on a multi-resolution analysis have been developed \cite{schroder:1995,sweldens:1996}.  Haar wavelets on the sphere for particular tessellation schemes have also been developed \cite{tenorio:1999,barreiro:2000}.
However, these discrete constructions may not lead to a stable basis.  Furthermore, the symmetry of the sphere is clearly lost when a tessellation scheme is adopted.
Other authors have focused on the continuous wavelet transform on the sphere.  
Continuous spherical wavelets have been constructed directly from spherical harmonic functions or Legendre polynomials by \cite{freeden:1997a,freeden:1997b}, however these solutions suffer from poor localisation and require an abstract dilation parameter that imposes a number of {\it ad hoc} assumptions.
A satisfactory extension of the continuous wavelet transform to the sphere is defined by \cite{holschneider:1996}, however this construction also requires an abstract dilation parameter and corresponding {\it ad hoc} assumptions. 
Others adopt a tangent bundle viewpoint \cite{torresani:1995,dahlke:1996}, thereby avoiding the necessity to define a dilation operator on the sphere.  
Many of these previous works also consider only azimuthally symmetric wavelets and do not make the extension to a directional analysis.
A consistent and satisfactory framework for a directional spherical wavelet transform has been
constructed and developed by 
\cite{antoine:1998,antoine:1999,antoine:2002,antoine:2004,bogdanova:2004,demanet:2003,wiaux:2005}.   This transform was originally derived entirely from group theoretic principles \cite{antoine:1998}, however the  formalism has since been reintroduced independently of the original group theoretic
formalism in an equivalent, practical and consistent approach \cite{wiaux:2005}.
A stereographic transform is adopted in these works to relate spherical wavelets and operators to their planar counterparts.
For a more detailed review of the attempts made to construct a wavelet transform on the sphere see \cite{antoine:1999,antoine:1998,mhaskar:2001}.

Recently, an alternative wavelet construction on the sphere has been derived by \cite{sanz:2006}.  In this setting the spherical dilation operator is defined by a scaling operation in spherical harmonic space.  This construction has the advantage that all \mbox{functions} and operations are defined directly on the sphere, rather than through projections to and from the plane.  \mbox{However}, the derivation presented by \cite{sanz:2006} is not fully general and is restricted to azimuthally symmetric wavelets.  We adopt the harmonic scaling idea for the dilation operator proposed by \cite{sanz:2006} but extend the spherical wavelet analysis to incorporate directional wavelets.  Directional wavelets are a powerful extension that allow one also to probe oriented structure in the analysed function (see \eg\ \cite{antoine:2004} for a discussion of directional wavelets).


The remainder of this paper is organised as follows.  We begin by outlining some mathematical preliminaries in \sectn{\ref{sec:prelim}}.  In \sectn{\ref{sec:analysis}} and \sectn{\ref{sec:synthesis}} we derive the spherical wavelet analysis and synthesis respectively.  In \sectn{\ref{sec:properties}} we show some properties of our spherical wavelet construction.  We demonstrate in \sectn{\ref{sec:construction}} the construction of spherical wavelets in our formalism.  Finally, concluding remarks are made in \sectn{\ref{sec:conclusions}}.

\section{Mathematical preliminaries}
\label{sec:prelim}

It is necessary first to outline some mathematical preliminaries before constructing the wavelet transform on the sphere.  By making all assumptions and definitions explicit we hope to avoid any
confusion over the conventions adopted.  The reader familiar with harmonic analysis on the sphere may skip this section and refer back to it as required.

 We consider the space of square integrable functions $L^2(\sphere,\dmu)$ on the unit 2-sphere \sphere, where \mbox{$\dmu(\sa) = \sin\saa \dx\saa \dx\sab$} is the usual rotation invariant measure on the sphere and $\sa \equiv (\sas) \in \sphere$ denotes spherical coordinates with colatitude \saa\ and longitude \sab.
A square integrable function on the sphere $f \in L^2(\sphere,\dmu)$ may be represented by the spherical harmonic expansion
\begin{displaymath}
f(\sa) = \sum_{\el=0}^\infty \sum_{\m=-\el}^\el \shc{f}{\el}{\m} \shfarg{\el}{\m}{\sa}
\spcend ,
\end{displaymath}
where the spherical harmonic coefficients are given by the usual projection on to the spherical harmonic basis functions
\begin{displaymath}
\shc{f}{\el}{\m} 
= 
\int_{\sphere}
\dmu(\sa) \:
f(\sa) \:
\shfargc{\el}{\m}{\sa} 
\spcend .
\end{displaymath}
The \cconj\ denotes complex conjugation.
We adopt the Condon-Shortley phase convention where the normalised 
spherical harmonics are defined by
\begin{displaymath}
\shfarg{\el}{\m}{\sa} = (-1)^\m \sqrt{\frac{2\el+1}{4\pi} 
\frac{(\el-\m)!}{(\el+\m)!}} \: 
\aleg{\el}{\m}{\cos\saa} \:
{\rm e}^{\img \m \sab}
\spcend ,
\end{displaymath}
where $\aleg{\el}{\m}{x}$ are the associated Legendre functions.  Using this normalisation the orthogonality of the spherical harmonic functions is given by
\begin{equation}
\label{eqn:shortho}
\int_\sphere
\dmu \:
\shfarg{\el}{\m}{\sa}
\shfargc{\el\p}{\m\p}{\sa}
=
\kron_{\el\el\p} \kron_{\m\m\p}
\spcend ,
\end{equation}
where $\delta_{ij}$ is Kronecker delta function.  


To construct a directional \cswt\ on the sphere one must consider rotations on the sphere.  Rotations on the sphere are characteristed  by the elements of the rotation group \sothree, which we parameterise in terms of the three Euler angles \mbox{$\rho=(\euls)$}.\footnote{We adopt the $zyz$ Euler convention corresponding to the rotation of a physical body in a \emph{fixed} co-ordinate system about the $z$, $y$ and $z$ axes by \eulc, \eulb\ and \eula\ respectively.}  The rotation of $f$ is defined by
\begin{displaymath}
[\rot(\eul) f](\sa) = f(\rho^{-1} \sa), \; \; \eul \in \sothree 
\spcend .
\end{displaymath}
It is also useful to characterise the rotation of a function on the sphere in harmonic space.  The rotation of a spherical harmonic basis function may be represented by a sum of weighted harmonics of the same \el\ \cite{brink:1993,ritchie:1999}:
\begin{equation}
\label{eqn:shrot:1}
\left[\rot(\eul)\shf{\el}{\m}\right](\sa) = 
\sum_{\n=-\el}^{\el} 
\dmatbig_{\n\m}^{\el}(\eul) \: \shfarg{\el}{\n}{\sa}
\spcend ,
\end{equation}
where the Wigner functions $\dmatbig_{\m\n}^{\el}(\eul)$ are described below.  From \eqn{\ref{eqn:shrot:1}} it is trivial to show that the harmonic coefficients of a rotated function are related to the coefficients of the original function by 
\begin{equation}
\label{eqn:shrot:2}
\left[\rot(\eul) f \right]_\elm = 
\sum_{\n=-\el}^{\el} 
\dmatbig_{\m\n}^{\el}(\eul) \: 
f_{\el\n}
\spcend .
\end{equation}
Note carefully the distinction between the indices of the Wigner functions in \eqn{\ref{eqn:shrot:1}} and \eqn{\ref{eqn:shrot:2}}.
The Wigner functions may be decomposed as \cite{brink:1993,ritchie:1999}
\begin{equation}
\label{eqn:d_decomp}
\dmatbig_{\m\n}^{\el}(\euls)
= {\rm e}^{-\img \m\eula} \:
\dmatsmall_{\m\n}^\el(\eulb) \:
{\rm e}^{-\img \n\eulc}
\spcend ,
\end{equation}
where the real polar \dmatsmall-matrix is defined by \cite{brink:1993}
\ifcol
  \begin{eqnarray*}
  \lefteqn{\dmatsmall^\el_{\m \n}(\eulb)  = 
  \sum_{t = \max(0, \m - \n)}^{\min(\el + \m, \el - \n)} (-1)^t} \nonumber\\
  && \times \frac{
  \left[(\el+\m)! \, (\el-\m)! \, (\el+\n)! \, (\el-\n)! \, \right]^{1/2}}
  {(\el+\m-t)! \, (\el-\n-t)! \, (t+\n-\m)! \, t!} \nonumber \\
  && \times \left[\cos\!\left(\frac{\eulb}{2} \right)\right]^{2\el+\m-\n-2t}
  \left[\sin \! \left(\frac{\eulb}{2} \right)\right]^{\n-\m+2t}
  \spcend ,
  \end{eqnarray*}
\else
  \begin{eqnarray*}
  \dmatsmall^\el_{\m \n}(\eulb) & = & 
  \sum_{t = \max(0, \m - \n)}^{\min(\el + \m, \el - \n)} (-1)^t
  \frac{
  \left[(\el+\m)! \, (\el-\m)! \, (\el+\n)! \, (\el-\n)! \, \right]^{1/2}}
  {(\el+\m-t)! \, (\el-\n-t)! \, (t+\n-\m)! \, t!} \nonumber \\
  && \times \left[\cos\!\left(\frac{\eulb}{2} \right)\right]^{2\el+\m-\n-2t}
  \left[\sin \! \left(\frac{\eulb}{2} \right)\right]^{\n-\m+2t}
  \spcend ,
  \end{eqnarray*}
\fi
and the sum over $t$ is defined so that the arguments of the factorials are non-negative.  
The Wigner functions satisfy the orthogonality condition
\begin{equation}
\label{eqn:wigner_ortho}
\int_\sothree 
\deul \:
\dmatbig_{\m\n}^{\el\cconj}(\eul) \:
\dmatbig_{\m\p\n\p}^{\el\p\cconj}(\eul)
=
\frac{8\pi^2}{2\el+1} \:
\kron_{\el\el\p} \kron_{\m\m\p} \kron_{\n\n\p}
\spcend ,
\end{equation}
where $\deul = \sin\eulb \dx \eula  \dx \eulb  \dx \eulc$.
Recursion formulae are available to compute rapidly the Wigner  \dmatsmall-matrices in the basis of either complex \cite{risbo:1996,choi:1999} or real \cite{ivanic:1996,blanco:1997} spherical harmonics.

\section{Analysis}
\label{sec:analysis}

We adopt the strategy suggested by \cite{sanz:2006} and define spherical wavelets and the spherical dilation operator in harmonic space.  We extend this idea here to a directional wavelet transform on the sphere.
The dilated wavelet $\wav(\sa; \scalea)$ is defined in harmonic space by
\begin{equation}
\label{eqn:wavgen}
\shc{\wav}{\el}{\m}(\scalea)
=
\sqrt{\frac{2\el+1}{8\pi^2}} \:
\wavm_\m(\el \scalea)
\spcend ,
\end{equation}
where $\shc{\wav}{\el}{\m}(\scalea$) are the spherical harmonic coefficients of $\wav(\sa; \scalea)$, $\wavm_\m(q)$ are the family of wavelet generating functions and $\scalea \in \realno^{+}_\ast$ is the real, strictly positive dilation parameter.  Note that the wavelet generating functions are not defined on the sphere but rather on the non-negative real line:
\mbox{$\wavm_\m\in L^2(\realno^+,\dx \xvect)$}, $\m \in \intno$ (although we consider only $|\m|<\el$ for $\shc{\wav}{\el}{\m}(\scalea$), we are in general free to define $\wavm_\m$, $\forall \m \in \intno$).  For the functions $\wav(\sa; \scalea)$ to classify as wavelets they must satisfy certain admissibility criteria so that the analysed function may be reconstructed perfectly from its wavelet coefficients.  We consider admissibility and the synthesis of a function from its wavelet coefficients in \sectn{\ref{sec:synthesis}}.

An overcomplete wavelet basis on the sphere may be constructed from the following spherical wavelet family:
\begin{displaymath}
\{
[\rot(\eul) \wav](\sa; \scalea)
\mid
\eul \in \sothree, \:
\scalea \in \realno^{+}_\ast
\}
\spcend .
\end{displaymath}
The directional \cswt\ of \mbox{$\sky \in L^2(\sphere,\dmu)$} is given by the projection onto each wavelet basis function in the usual manner:
\begin{equation}
\label{eqn:cswt}
\skywav(\scalea, \eul) 
\equiv
\int_{\sphere}
\dmu(\sa) \:
\sky(\sa) \:
[\rot(\eul) \wav]^\cconj(\sa; \scalea)
\spcend .
\end{equation}

The transform is general in the sense that all orientations in the
rotation group \sothree\ are considered, thus directional structure is
naturally incorporated.  It is important to note, however, that only
\emph{local} directions make any sense on \sphere.  There is no global
way of defining directions on the sphere\footnote{There is no
differentiable vector field of constant norm on the sphere and hence
no global way of defining directions.} -- there will always be some
singular point where the definition fails.  

It is useful to represent the \cswt\ in terms of harmonic coefficients.
Substituting into \eqn{\ref{eqn:cswt}} the spherical harmonic expansion of the analysed function and of the rotated wavelet, and noting the orthogonality of the spherical harmonics given by \eqn{\ref{eqn:shortho}}, one may represent the \cswt\ by
\begin{equation}
\label{eqn:cswt_harmonic}
\skywav(\scalea, \eul) 
=
\sumlmn
\shc{\sky}{\el}{\m} \:
\Dlmnpc \:
\shcc{\wav}{\el}{\n}(\scalea)
\spcend . 
\end{equation}

\section{Synthesis}
\label{sec:synthesis}

For the multi-resolution analysis we describe in the previous section to classify as a wavelet analysis, it must be possible to reconstruct perfectly the original function from its wavelet coefficients and the wavelet basis functions.  To ensure that this is possible the wavelets must satisfy certain admissibility criteria.  We derive the admissibility criteria and perfect reconstruction formula in this section.  We then relate the admissibility condition for each wavelet basis function to an equivalent condition for the family of wavelet generating functions.

Consider the expression
\begin{equation}
\label{eqn:recon:1}
\int_0^\infty \frac{\dx\scalea}{\scalea} \:
\int_{\sothree} \dx\eul \:
\skywav(\scalea, \eul) \:
[\rot(\eul) L_\wavm \wav](\sa; \scalea)
\spcend ,
\end{equation}
where the $L_\wavm$ operator in $L^2(\sphere,\dmu)$ is defined by the
action
\begin{equation}
\shc{(L_\wavm g)}{\el}{\m} = \shc{g}_\elm / \admissC
\end{equation}
on the spherical harmonic coefficients of functions \mbox{$g \in L^2(\sphere,\dmu)$} ($\admissC$ is defined below).
Substituting the harmonic representation of the wavelet coefficients given by \eqn{\ref{eqn:cswt_harmonic}} and the harmonic representation of $[\rot(\eul) L_\wavm \wav](\sa; \scalea)$ into \eqn{\ref{eqn:recon:1}}, and noting the orthogonality of the Wigner functions defined by \eqn{\ref{eqn:wigner_ortho}}, \eqn{\ref{eqn:recon:1}} may be rewritten as
\ifcol
  \begin{eqnarray}
  \label{eqn:recon:2}
  &\displaystyle  
  \sumlm
  \shc{\sky}{\el}{\m} \:
  \shfarg{\el}{\m}{\sa} \:
  \frac{1}{\admissC} \nonumber \\ &\displaystyle \times \:
  \frac{8\pi^2}{2\el+1}
  \sum_{\n=-\el}^{\el}
  \int_0^\infty \frac{\dx\scalea}{\scalea} \:
  | \shc{\wav}{\el}{\n}(\scalea) |^2
  \spcend .
  \end{eqnarray}
\else
  \begin{equation}
  \label{eqn:recon:2}
  \sumlm
  \shc{\sky}{\el}{\m} \:
  \shfarg{\el}{\m}{\sa} \:
  \frac{1}{\admissC} \:
  \frac{8\pi^2}{2\el+1}
  \sum_{\n=-\el}^{\el}
  \int_0^\infty \frac{\dx\scalea}{\scalea} \:
  | \shc{\wav}{\el}{\n}(\scalea) |^2
  \spcend .
  \end{equation}
\fi
For perfect reconstruction we require the admissibility criteria
\begin{equation}
\label{eqn:admiss}
0 <
\admissC \equiv
\frac{8\pi^2}{2\el+1}
\sum_{\m=-\el}^{\el}
\int_0^\infty \frac{\dx\scalea}{\scalea} \: 
| \shc{\wav}{\el}{\m}(\scalea) |^2
< \infty
\spcend ,
\end{equation}
to hold $\forall \el$.  We call \admissC\ the admissibility constant.
If admissibility is satisfied it is apparent from \eqn{\ref{eqn:recon:2}} that the original function may be reconstructed by
\ifcol
  \begin{eqnarray}
  & \displaystyle
  \sky(\sa) = 
  \int_0^\infty \frac{\dx\scalea}{\scalea} \:
  \int_{\sothree} \dx\eul \:
  \skywav(\scalea, \eul) \nonumber \\ & \displaystyle \times \:
  [\rot(\eul) L_\wavm \wav](\sa; \scalea)
  \spcend .
  \end{eqnarray}
\else
  \begin{equation}
  \sky(\sa) = 
  \int_0^\infty \frac{\dx\scalea}{\scalea} \:
  \int_{\sothree} \dx\eul \:
  \skywav(\scalea, \eul) \: 
  [\rot(\eul) L_\wavm \wav](\sa; \scalea)
  \spcend .
  \end{equation}
\fi
We relate the admissibility criteria that each spherical wavelet must satisfy to an equivalent admissibility condition for the family of wavelet generating functions.  Substituting \eqn{\ref{eqn:wavgen}} into \eqn{\ref{eqn:admiss}} and making the change of variables $q=\el \scalea$, the admissibilty constant may be rewritten as
\begin{equation}
\label{eqn:admiss:2}
0 <
\admissC =
\sum_{\m=-\el}^{\el}
\int_0^\infty \frac{\dx q}{q} \: 
| \wavm_\m(q) |^2
< \infty
\spcend .
\end{equation}
Since $\int_0^\infty \frac{\dx q}{q} \: | \wavm_\n(q) |^2$ is always non-negative it is possible to recast the admissibility condition on the family $\wavm_\m$ in the following form:
$
\int_0^\infty \frac{\dx q}{q} \: 
| \wavm_\m(q) |^2 < \infty
$, 
$\forall \m \in \intno$
and 
$\exists \m \in \intno$ such that
$
\int_0^\infty \frac{\dx q}{q} \: 
| \wavm_\m(q) |^2 \neq 0
$
.

\section{Properties}
\label{sec:properties}

In this section we describe some properties of the wavelet basis functions and the wavelet coefficients of the \cswt\ derived in the previous sections.

\subsection{Compensation}

Admissible wavelets are compensated:
\begin{displaymath}
\int_{\sphere}
\dmu(\sa) \:
\wav(\sa; \scalea)
=0, \:
\forall \scalea \in \realno^+_\ast
\spcend .
\end{displaymath}
We show this property by noting that admissibility imposes the condition $\wavm_\m(0) = 0$, $\forall \m \in \intno$.  
Considering the spherical harmonic transform of the wavelet,
\begin{displaymath}
\sqrt{\frac{2\el+1}{8\pi^2}} \:
\wavm_\m(\el\scalea)
=
\int_{\sphere}
\dmu(\sa) \:
\wav(\sa; \scalea) \:
\shfargc{\el}{\m}{\sa}
\spcend ,
\end{displaymath}
and setting $\el=0$ and $\m=0$ we obtain the compensation property.

\subsection{Wavelet energy}

The energy of a wavelet at a particular dilation scale is given by 
\begin{eqnarray}
\| \wav(\scalea) \|^2
&\equiv& 
\int_{\sphere}
\dmu(\sa) \:
| \wav(\sa; \scalea) |^2 \nonumber \\
&=& 
\sumlm 
| \shc{\wav}{\el}{\m}(\scalea) |^2 \nonumber \\
&=&
\sumlm 
\frac{2\el+1}{8\pi^2} \:
| \wavm_{\m}(\el\scalea) |^2 \nonumber
\spcend ,
\end{eqnarray}
where in the second line we have made use of the orthogonality of the spherical harmonics described by \eqn{\ref{eqn:shortho}}.

\subsection{Function energy}

The energy of the analysed function is given by
\begin{eqnarray}
\| \sky \|^2
&\equiv& 
\int_{\sphere}
\dmu(\sa) \:
| \sky(\sa) |^2 \nonumber \\
&=& 
\sumlm 
| \shc{\sky}{\el}{\m} |^2 \nonumber
\spcend ,
\end{eqnarray}
where in the second line we have made use of the orthogonality of the spherical harmonics described by \eqn{\ref{eqn:shortho}}.  We may also relate the energy of the function directly to its spherical wavelet coefficients:
\begin{displaymath}
\| \sky \|^2
= 
\int_0^\infty \frac{\dx\scalea}{\scalea} \:
\int_{\sothree} \dx\eul \:
\skyLwav(\scalea, \eul)  \:
\skywav{}^\cconj(\scalea, \eul) \:
\spcend .
\end{displaymath}
To show this property we substitute the harmonic representation of the wavelet coefficients given by \eqn{\ref{eqn:cswt_harmonic}} and note the {ortho\-gonality} of both the spherical harmonics and Wigner functions described by \eqn{\ref{eqn:shortho}} and \eqn{\ref{eqn:wigner_ortho}} respectively.

\subsection{Wavelet domain variance}

The variance of the wavelet coefficients for a particular scale is used by \cite{sanz:2006} as a measure of the concentration properties of the wavelet transform.  However, this measures the variance of the wavelet coefficient {\it values} and {\it not} the localisation of energy in wavelet space.  In any case, it is useful to relate variance in the wavelet domain to the energy of the analysed function and wavelet.

The variance of the wavelet coefficients for a particular scale is given by
\begin{displaymath}
  \left[ \Delta \skywav(\scalea) \right]^2 =
  \left< \left| \skywav(\scalea,\eul) \right|^2 \right> 
  - \left| \left< \skywav(\scalea,\eul) \right> \right|^2
  \spcend ,
\end{displaymath}
where we define the expected value over rotations of \mbox{$G(\eul) \in L^2(\sothree, \dx\eul)$} by
\begin{displaymath}
  \left< G(\eul) \right> = \frac{1}{8\pi^2} \int_\sothree \dx\eul \: G(\eul)
  \spcend .
\end{displaymath}
The expected value of the wavelet coefficients is given by
\begin{displaymath}
  \left< \skywav(\scalea,\eul) \right>
  = 
  \sumlmn
  \shc{\sky}{\el}{\m} \:
  \shcc{\wav}{\el}{\n}(\scalea) \:
  \left< \Dlmnpc \right>
\end{displaymath}
and
\begin{eqnarray*}
  \left< \Dlmnpc \right> &=&
  \frac{1}{8\pi^2}
  \int_0^{2\pi} \dx \eula \: {\rm e}^{\img \m \eula}
  \int_0^{\pi} \dx \eulb \: \sin\eulb \: \dmatsmall_{\m\n}^{\el\cconj}(\eulb) \nonumber \\
  &&\times
  \int_0^{2\pi} \dx \eulc \: {\rm e}^{\img \n \eulc} \nonumber \\
  &=& \frac{1}{2} \, \kron_{\m 0} \, \kron_{\n 0} \int_{-1}^{1} \leg{\el}{x} \dx x \nonumber \\
  &=& \kron_{\el 0} \, \kron_{\m 0} \, \kron_{\n 0}
  \spcend ,
\end{eqnarray*}
where we have applied the Wigner \dmatbig-function decomposition described by \eqn{\ref{eqn:d_decomp}}.  The expected value of the wavelet coefficients is thus
\begin{displaymath}
  \left< \skywav(\scalea,\eul) \right>
  = 
  \shc{\sky}{0}{0} \:
  \shcc{\wav}{0}{0}(\scalea),
\end{displaymath}
which is zero for an admissible wavelet.
The variance of the wavelet coefficients is therefore given by
\begin{eqnarray}
\left[ \Delta \skywav(\scalea) \right]^2 &=&
  \left< \left| \skywav(\scalea,\eul) \right|^2 \right> \nonumber \\
  &=& 
  \sumlmn
  \frac{1}{2\el+1} \:
  | \shc{\sky}{\el}{\m} |^2 \:
  | \shc{\wav}{\el}{\n}(\scalea) |^2
  , \nonumber
\end{eqnarray}
where the second line follows from the harmonic representation of the wavelet coefficients given by \eqn{\ref{eqn:cswt_harmonic}} and by the orthogonality of the Wigner functions described by \eqn{\ref{eqn:wigner_ortho}}.

\section{Wavelet construction}
\label{sec:construction}

In the spherical wavelet framework outlined herein wavelets on the sphere are constructed directly in spherical harmonic space.  Spherical wavelets may be constructed in harmonic space from the analogue of the Fourier definition of planar Euclidean wavelets.  We must then check that the candidate wavelets are admissible. In this section we demonstrate the construction of a directional spherical wavelet with an example.

We construct the spherical analogue of the Morlet wavelet using our spherical wavelet formalism.  Note that this differs from the definition of the spherical Morlet wavelet constructed from the projection of the planar Morlet wavelet on to the sphere \cite{antoine:2002,antoine:2004}.  The planar Morlet wavelet is defined by a Gaussian in Fourier space centered on the wave vector of the wavelet, hence we define the analogue of the Morlet wavelet on the sphere by a Gaussian in spherical harmonic space.  Various associations between Fouier and spherical harmonic space are possible.  Here we choose to assocate $\el$ and $\m$ in spherical harmonic space with the $x$ and $y$ components of a vector in Fourier space respectively.\footnote{
Alternatively one may choose to associate \el\ with the length of a vector in Fourier space and \m\ with its phase.  We have also considered this association, although the support of the resulting functions is not so compact.  Moreover, we have not yet shown that the candidate wavelets that result are admissible. 
}
The wavelet generating functions are defined by
\begin{equation}
\wavm_\m(\el\scalea) = 
{\rm e}^{-\frac{(\el\scalea-L)^2 + (\m-M)^2}{2}}
-
{\rm e}^{-\frac{(\el\scalea)^2 + L^2 + (\m-M)^2}{2}}
\spcend ,
\end{equation}
where $L\in\natno$ and $M\in\intno$, $|M|<L$ define the centre of the Gaussian when $\scalea=1$.  The correction term subtracted is included to ensure admissibility (and takes a similar form to the correction used to ensure the planar Morlet wavelet is admissible \cite{antoine:2004}).  For numerical purposes, when $L\geq5$ the correction is numerically negligible and may simply be ignored.  Moreover, for numerical purposes it is possible to relate the maximum spherical harmonic \elmax\ that must be considered to $L$ and \scalea:
$\elmax = \lceil (L + \sqrt{-2 \: {\rm ln} \varepsilon})/\scalea \rceil$, 
where $\varepsilon$ is the allowable relative error and $\lceil x \rceil$ is the smallest integer greater than or equal to $x$.  We adopt the relation \mbox{$\elmax = \lceil (L+5)/\scalea \rceil$}, which ensures $\varepsilon<10^{-5}$.

Before proceeding it is necessary to check that the candidate wavelet generating functions generate admissible spherical wavelets.  We must show that he admissibility integral for the wavelet generating functions, given by
\begin{displaymath}
\int_0^\infty \frac{\dx q}{q} \: 
|\wavm_\m(q)|^2
=
{\rm e}^{-[L^2+(m-M)^2]}
\int_0^\infty \frac{\dx q}{q} \: 
{\rm e}^{-q^2}
({\rm e}^{qL} - 1)^2
\spcend .
\end{displaymath}
converges.  To do this, we split the range of integration up into
the interval $(0,\epsilon)$, where $\epsilon>0$ is small, and
the interval $(\epsilon,\infty)$. For the latter, we can appeal to the
comparison test for integrals since the ratio of $\frac{1}{q}
{\rm e}^{-q^2} ({\rm e}^{qL}-1)^2 $ to $q^{-2}$ tends to zero as $q \rightarrow
\infty$, for any $L$, and $\int_{\epsilon}^{\infty} \frac{dq}{q^2}$ is
finite (provided $\epsilon>0$). For the lower range, we note that the
expansion of the integrand for small $q$ is $q L^2$, and
$\int_0^{\epsilon} dq \, q L^2$ is finite.

In this section we have demonstrated the construction of admissible wavelets on the sphere in the spherical wavelet framework described herein.  In particular, we have constructed the analogue of the Morlet planar wavelet.  In \fig{\ref{fig:wavelets_r1}} and \fig{\ref{fig:wavelets_r0p5}} we plot this spherical wavelet for $(L,M)=(10,0)$ for a dilation scale of $\scalea=1$ and $\scalea=0.5$ respectively.  
Note that this construction yields a spherical wavelet that is quite different to the spherical Morlet wavelet constructed by projecting the planar Morlet wavelet on to the sphere (to make a comparison see \cite{antoine:2002,antoine:2004}).
Other wavelets on the sphere may be constructed in our framework in a similar manner.
Although we have constructed complex spherical wavelets, real wavelets may be constructed simply by ensuring the reality condition 
$\wavm_{-\m}(q) = (-1)^\m \wavmc_{\m}(q)$
is satisfied.

\newlength{\sparawidth}
\newlength{\scolourwidth}
\setlength{\sparawidth}{40mm}
\setlength{\scolourwidth}{40mm}

\begin{figure}
\centering
\mbox{
  \subfigure[$\zreal \{ \wav(\sa; \scalea=1) \} $]{\includegraphics[viewport= 350 0 900 850,width=\sparawidth,clip]{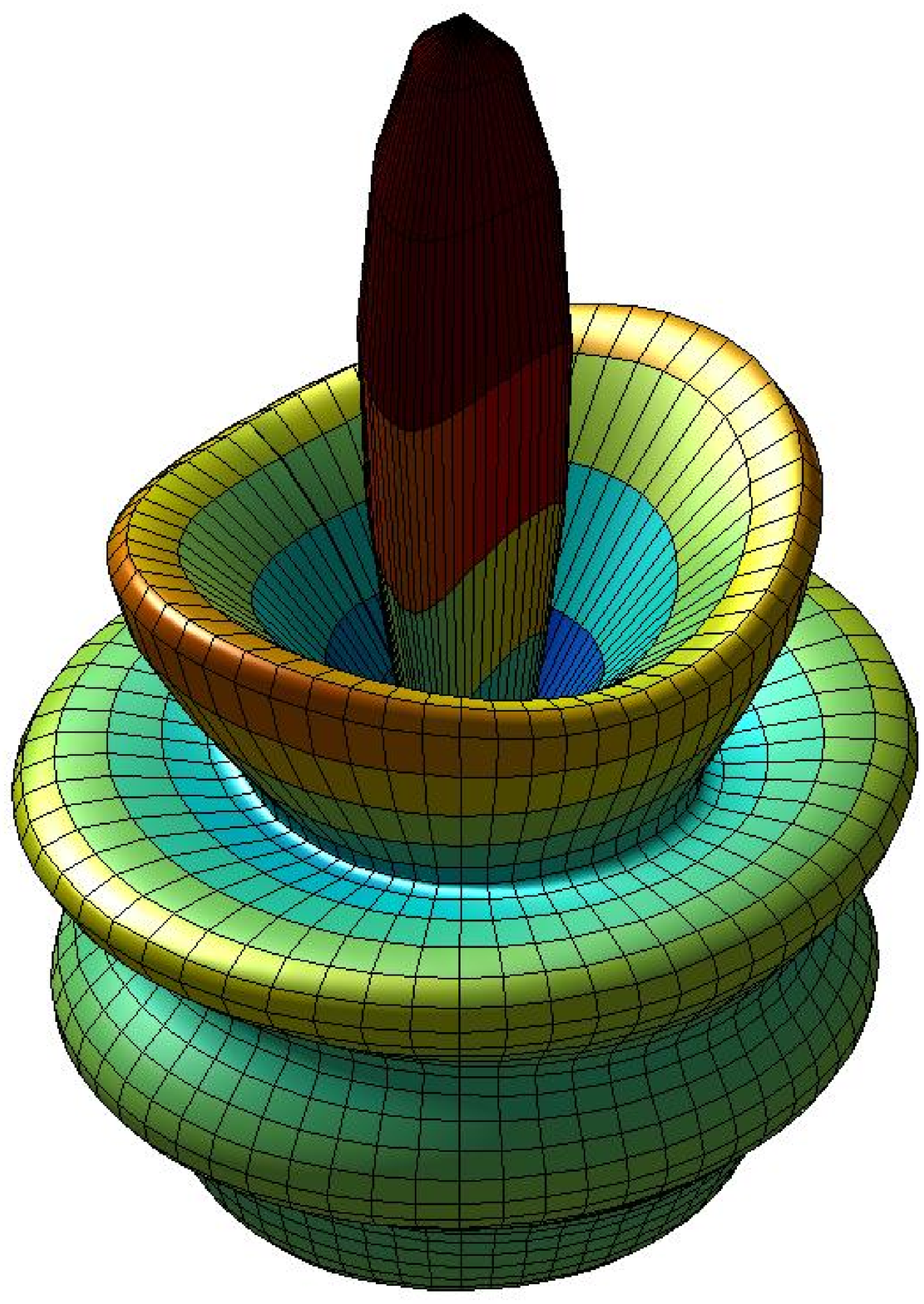}
  \includegraphics[viewport= 250 0 1000 850,width=\scolourwidth,clip]{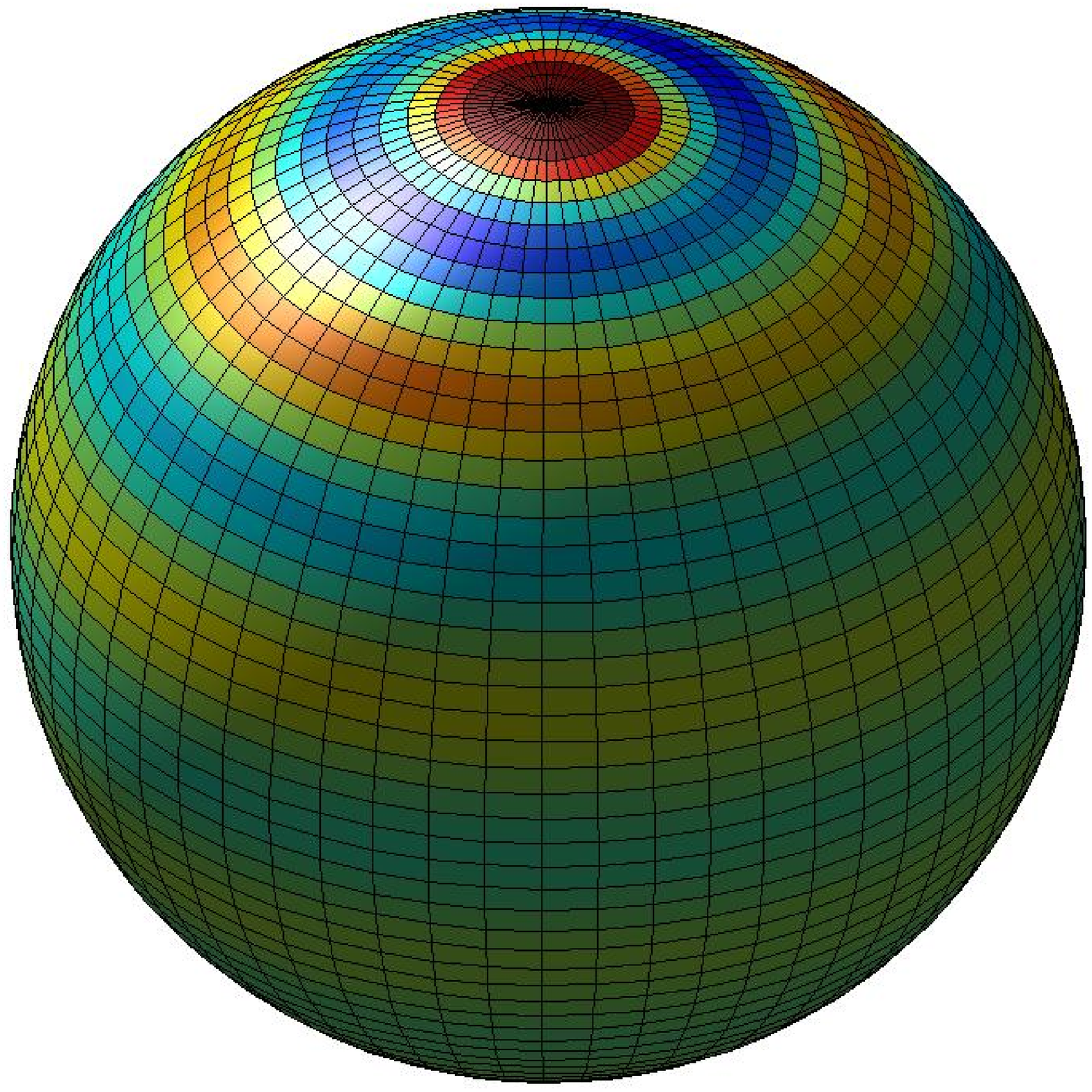}} } \quad
\mbox{
  \subfigure[$\zimag \{ \wav(\sa; \scalea=1) \} $]{\includegraphics[viewport= 320 0 920 850,width=\sparawidth,clip]{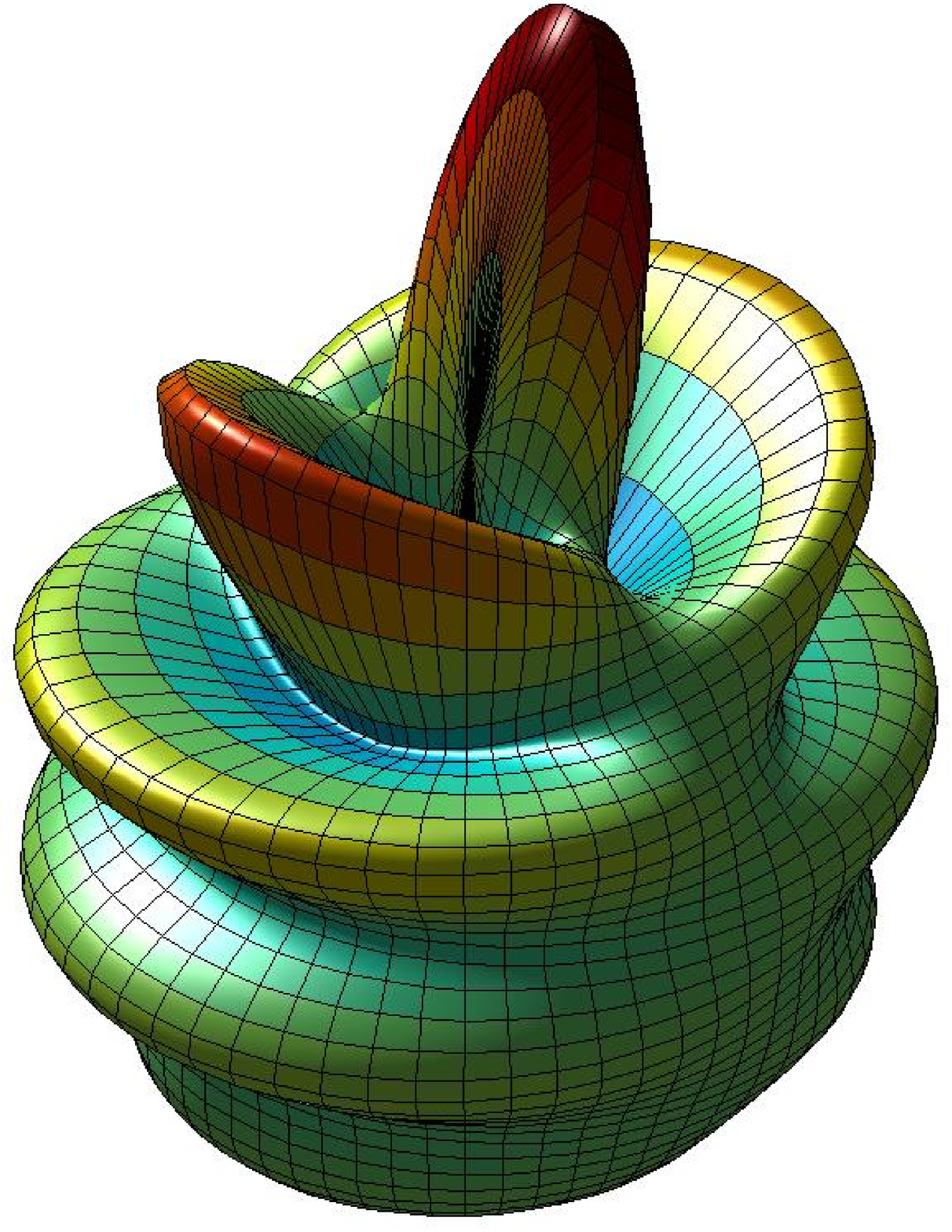}
  \includegraphics[viewport= 250 0 1000 850,width=\scolourwidth,clip]{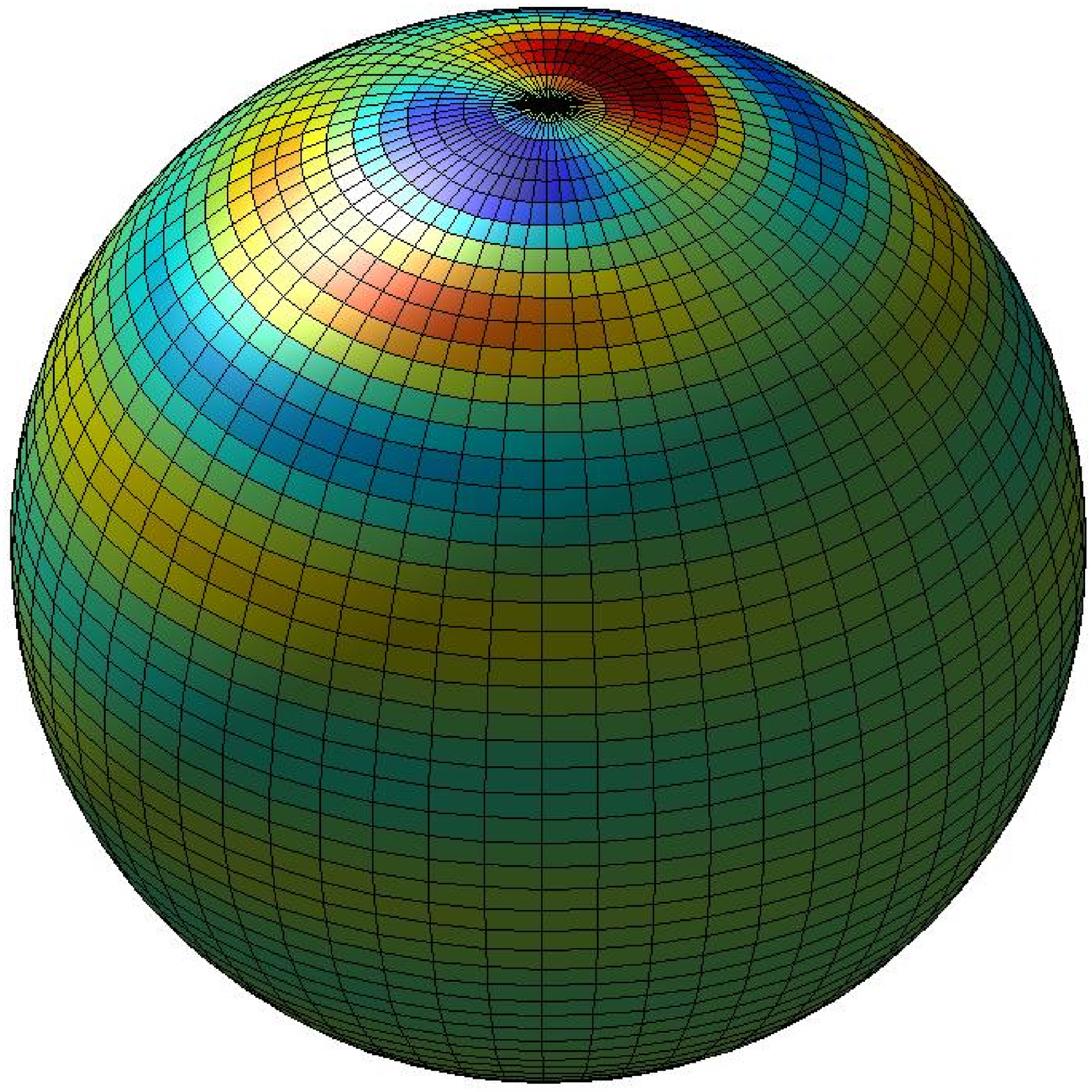}} } \quad
\mbox{
  \subfigure[$| \wav(\sa; \scalea=1) | $]{\includegraphics[viewport= 350 0 900 850,width=\sparawidth,clip]{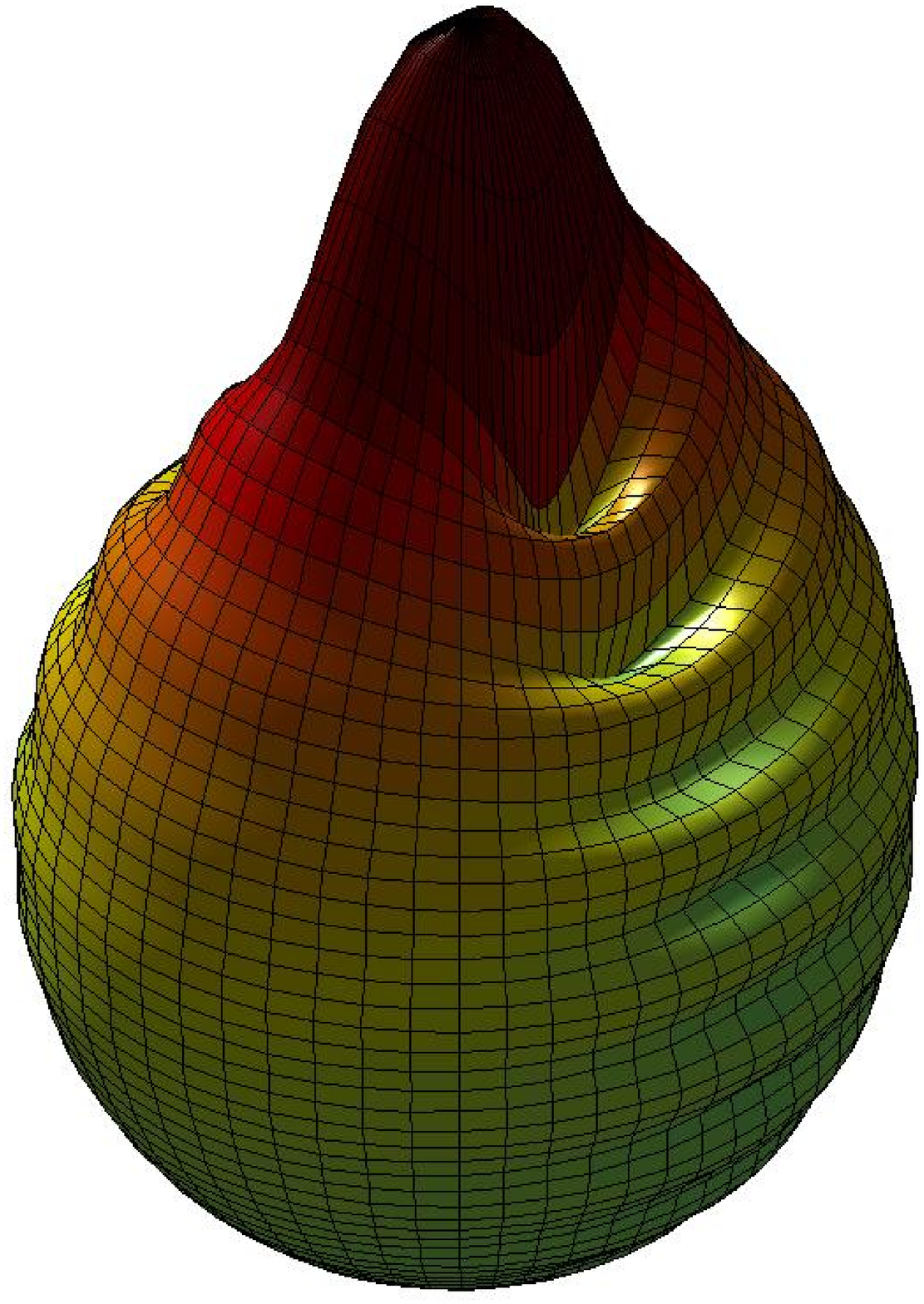}
  \includegraphics[viewport= 250 0 1000 850,width=\scolourwidth,clip]{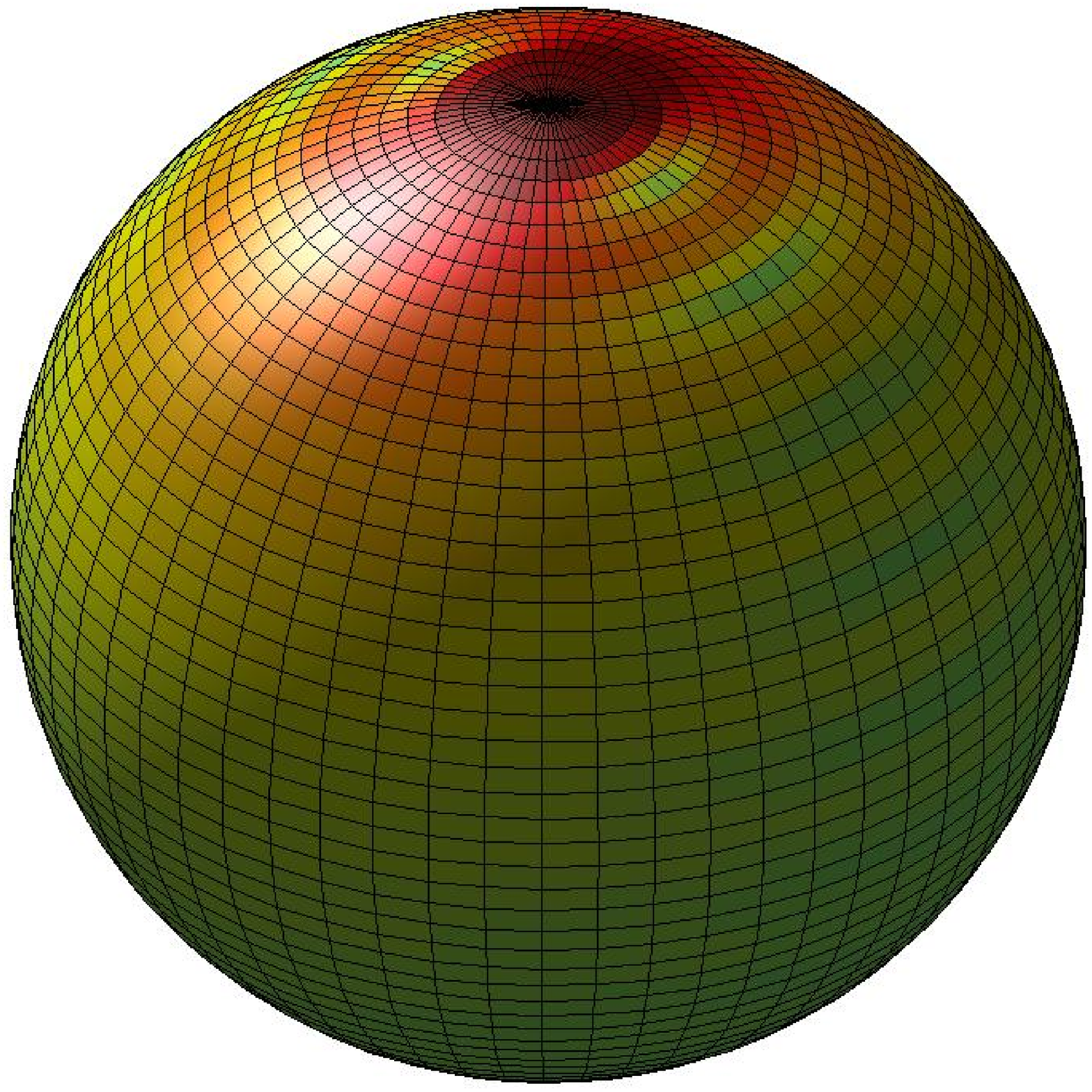}} }

\caption{Plots of the analogue of the Morlet wavelet constructed on the sphere for $\scalea=1$ using the formalism outlined herein.  Parametric plots are shown in the left column of panels; colour plots on the sphere are shown in the right column of panels.}
\label{fig:wavelets_r1}
\end{figure}

\begin{figure}
\centering
\mbox{
  \subfigure[$\zreal \{ \wav(\sa; \scalea=0.5) \} $]{\includegraphics[viewport= 350 0 900 850,width=\sparawidth,clip]{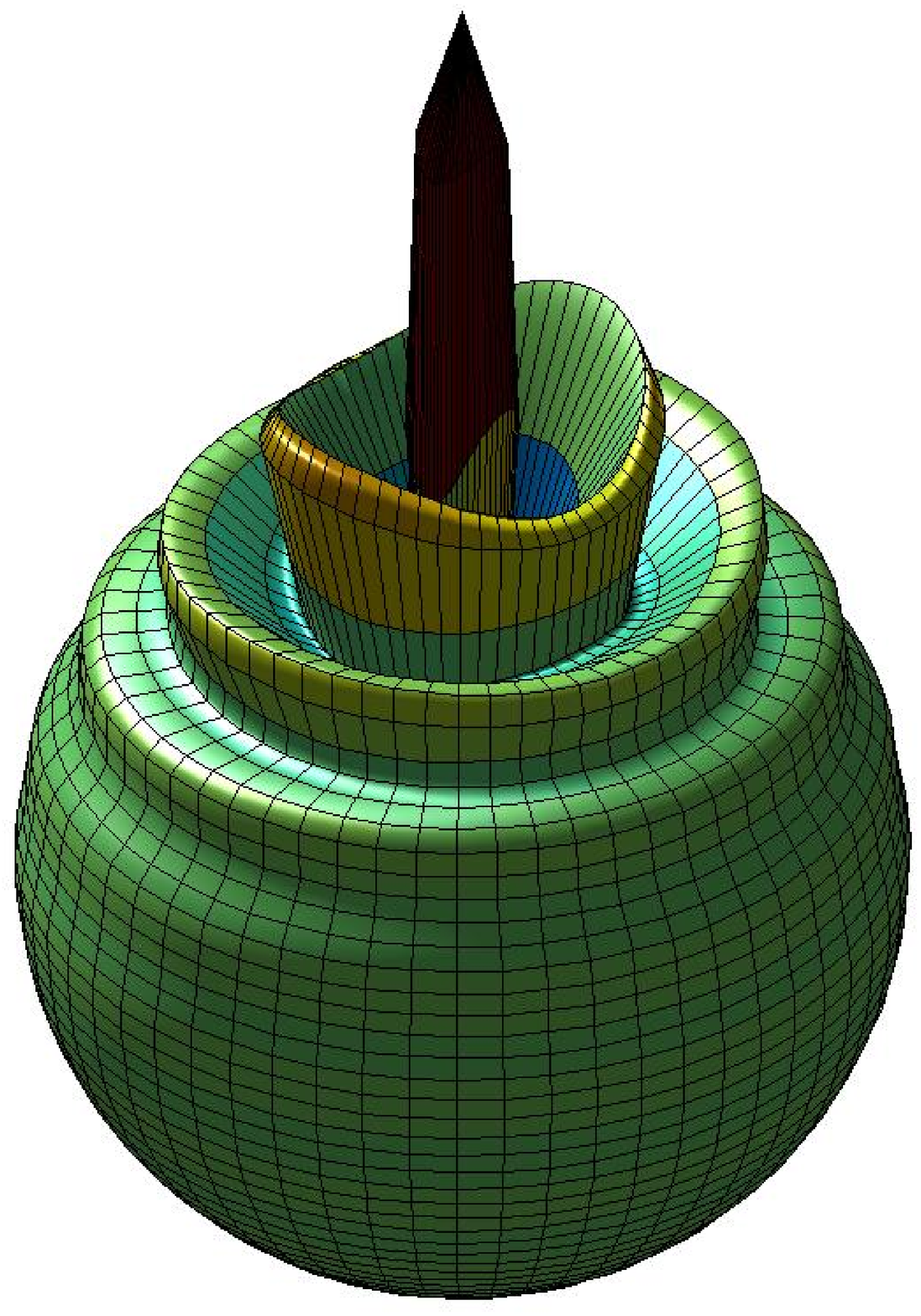}
  \includegraphics[viewport= 250 0 1000 850,width=\scolourwidth,clip]{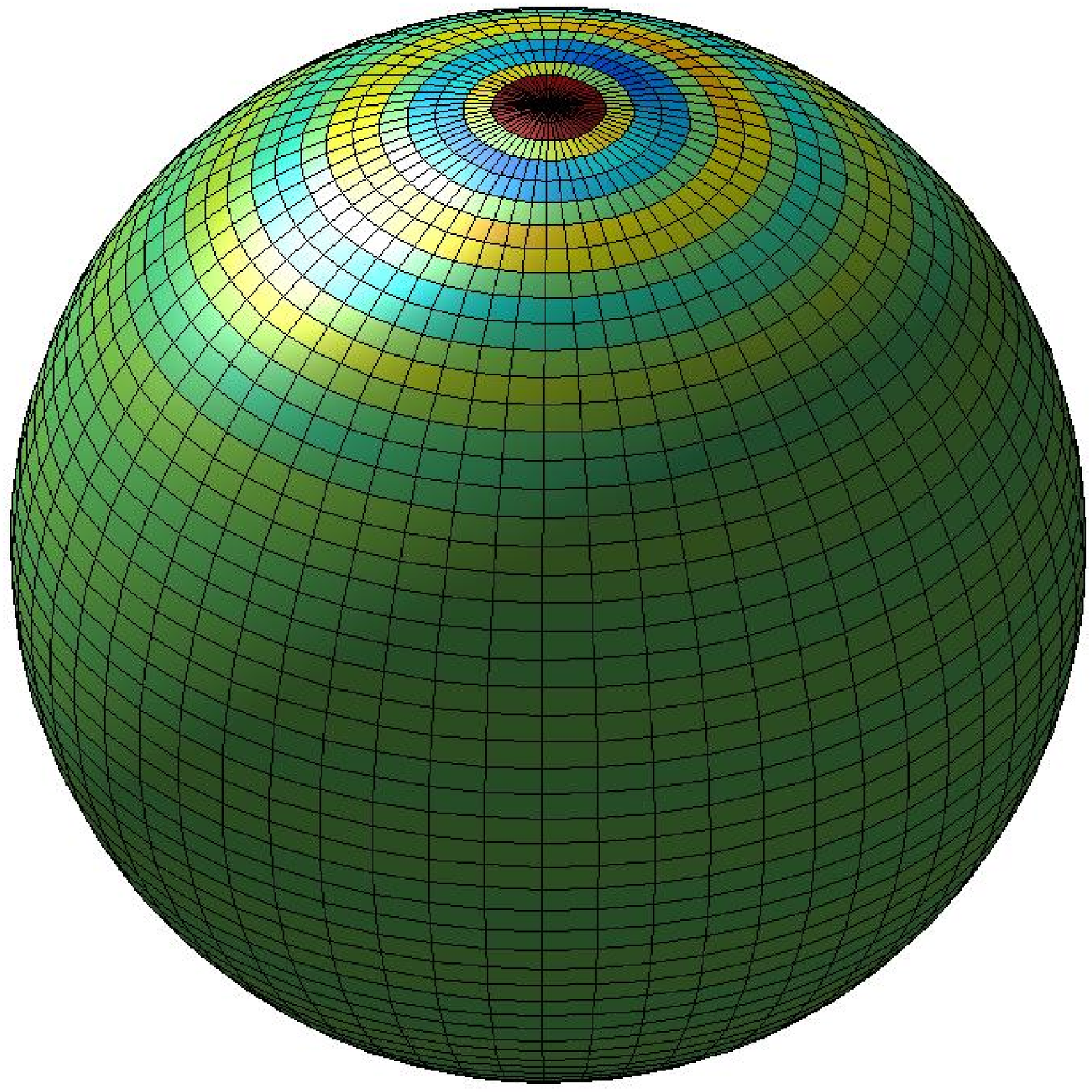}} } \quad
\mbox{
  \subfigure[$\zimag \{ \wav(\sa; \scalea=0.5) \} $]{\includegraphics[viewport= 320 0 920 850,width=\sparawidth,clip]{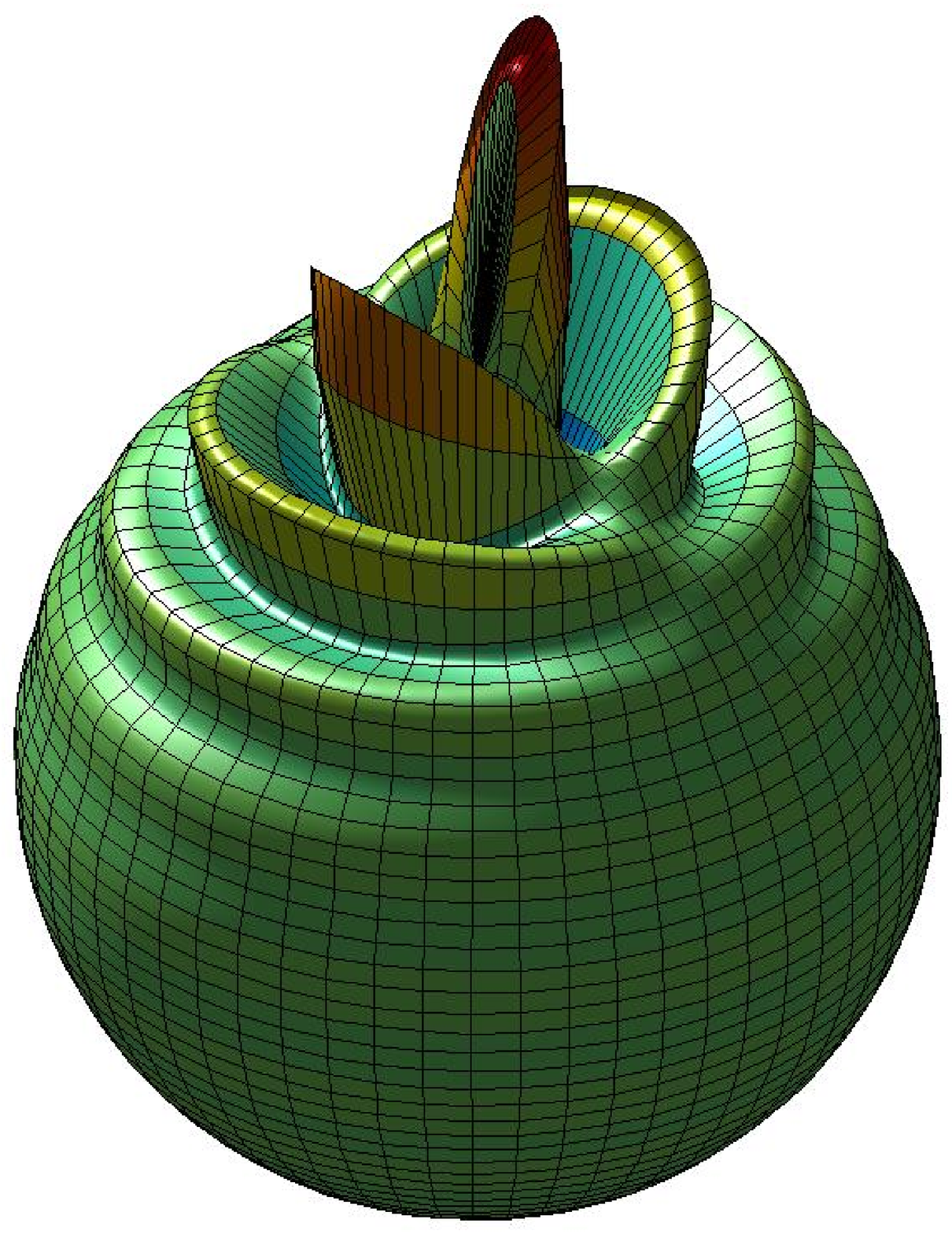}
  \includegraphics[viewport= 250 0 1000 850,width=\scolourwidth,clip]{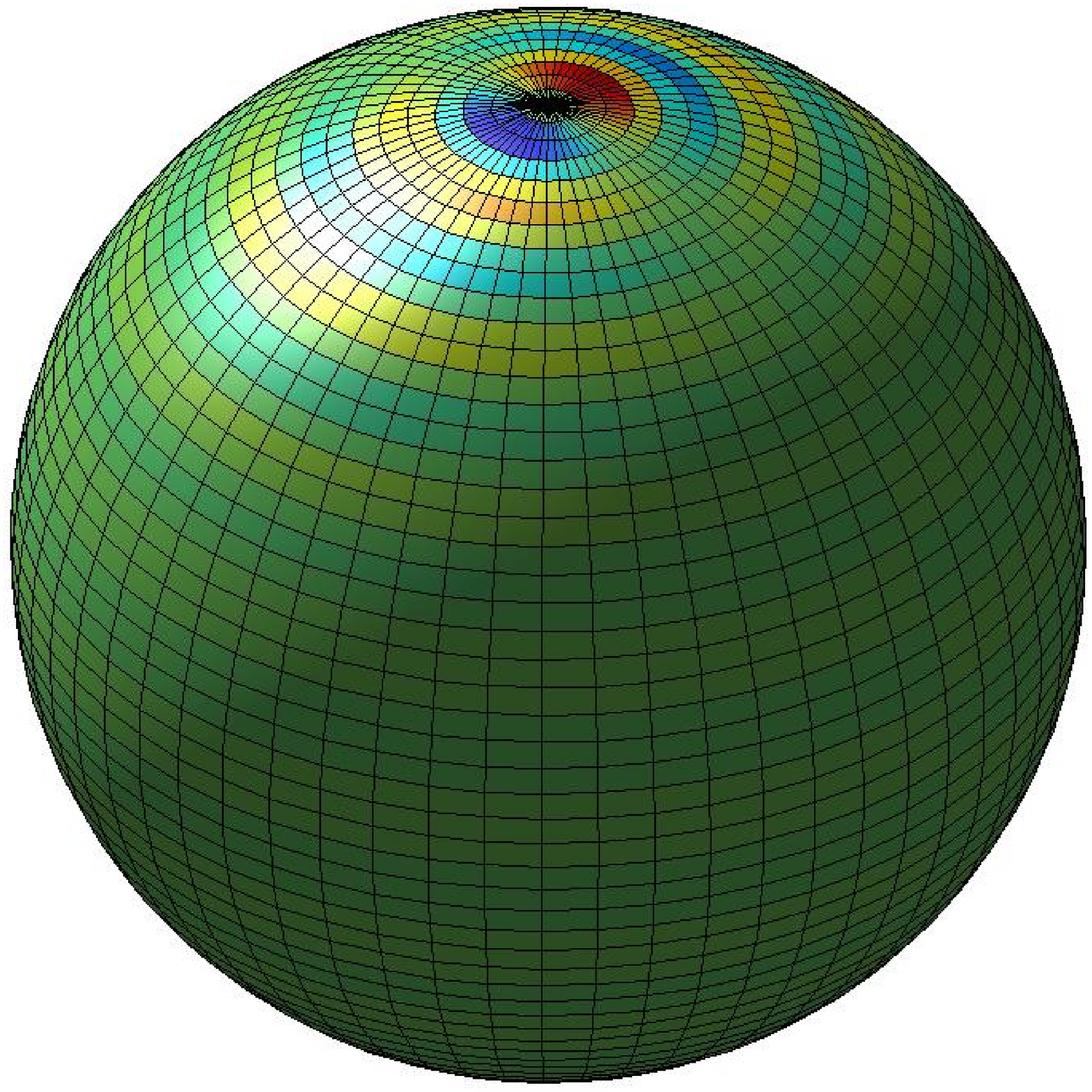}} } \quad
\mbox{
  \subfigure[$| \wav(\sa; \scalea=0.5) | $]{\includegraphics[viewport= 350 0 900 850,width=\sparawidth,clip]{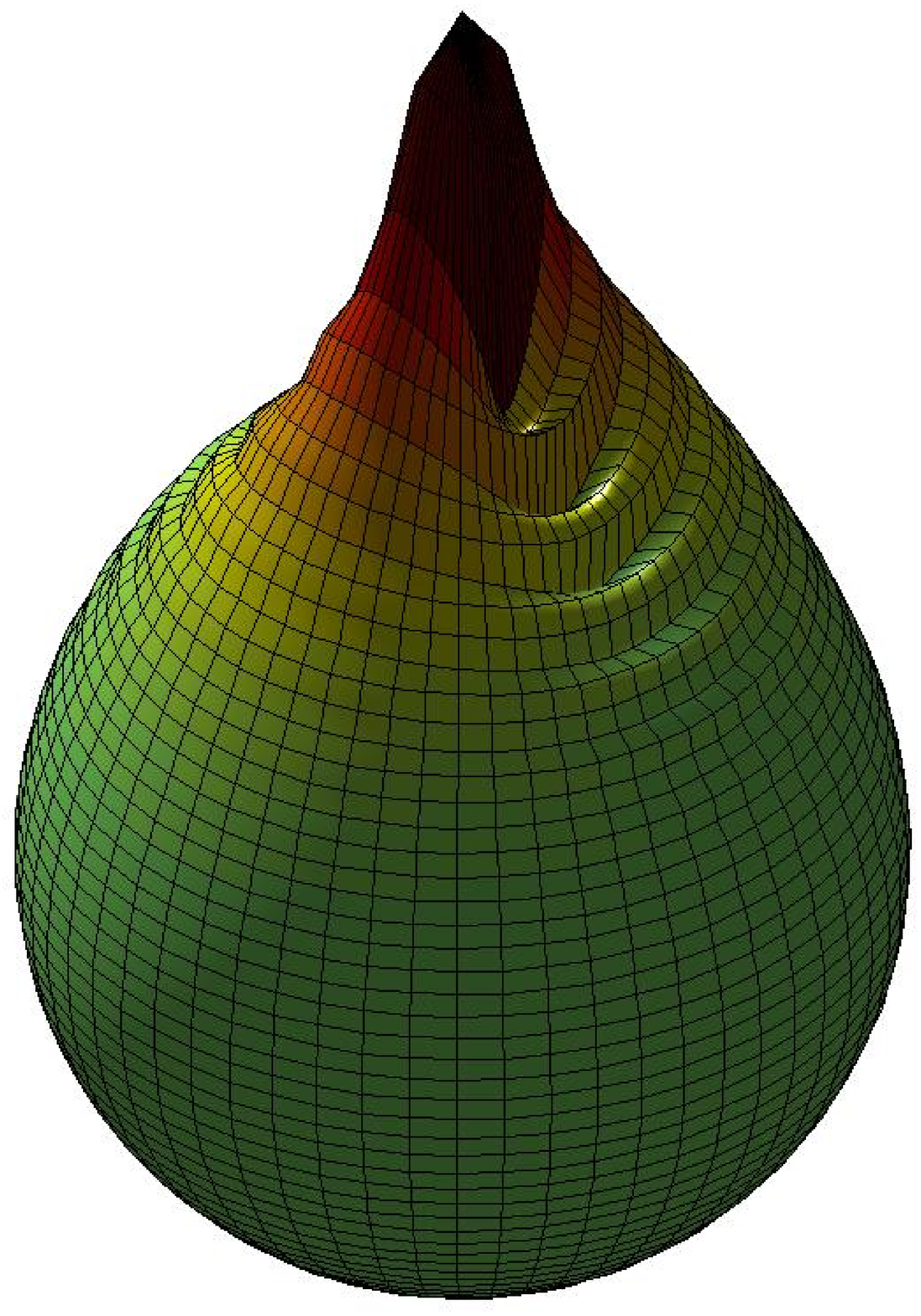}
  \includegraphics[viewport= 250 0 1000 850,width=\scolourwidth,clip]{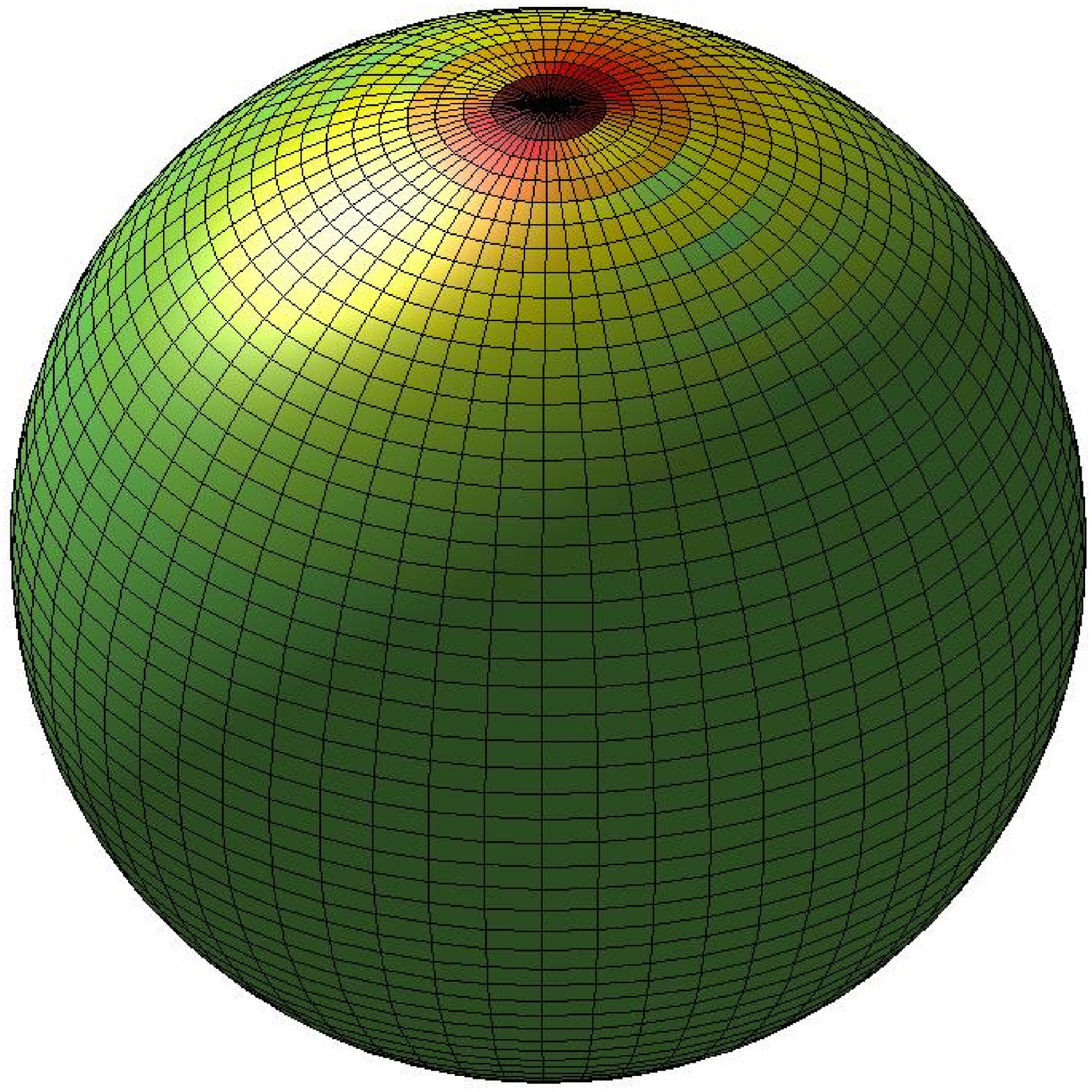}} }

\caption{Plots of the analogue of the Morlet wavelet constructed on the sphere for $\scalea=0.5$ using the formalism outlined herein.  Parametric plots are shown in the left column of panels; colour plots on the sphere are shown in the right column of panels. For illustration purposes the colour map and distance of the parametric plots from the unit sphere have been scaled by one fifth in this figure relative to the plots for $\scalea=1$.
}
\label{fig:wavelets_r0p5}
\end{figure}

\section{Concluding remarks}
\label{sec:conclusions}

We have presented the construction of a new directional continuous wavelet transform on the sphere.  Dilation on the sphere is performed by a scaling in harmonic space, as recently proposed by \cite{sanz:2006}.  We extend the spherical wavelet analysis presented by \cite{sanz:2006} to a more general directional framework, allowing one to probe oriented structure of the analysed function.  Our framework has the advantage that all wavelets and operators are defined directly on the sphere.  Once the wavelet analysis is performed, the original function may be reconstructed perfectly from its wavelet coefficients and the wavelet basis functions.  We have proved the corresponding synthesis formula and have derived the resulting admissibility criteria that the wavelets must satisfy.  Wavelets are constructed in our framework in spherical harmonic space.  We have demonstrated the construction of a wavelet basis with the example of the analogue of the Morlet wavelet in our spherical wavelet setting.  Other wavelets on the sphere may be constructed in our framework in a similar manner.

Not only is the extension of a wavelet analysis to a spherical manifold of important theoretical interest, it is also of important practical use in order to extend the potential benefits afforded by a wavelet analysis to data that are defined on the sphere.  For data sets of practical size, that may contain tens of millions of pixels, it is imperative to have a fast algorithm for performing the spherical wavelet analysis.  In a recent work we have presented fast algorithms to perform a directional \cswt\ \cite{mcewen:2006a} (for the \cswt\ derived by \cite{antoine:1998,antoine:1999,antoine:2002,antoine:2004,bogdanova:2004,demanet:2003,wiaux:2005}).  Fortunately, for a given dilation the analysis formula described herein is identical to that considered in these works.  Our fast algorithms are therefore directly applicable to our new directional wavelet construction.

The application of our spherical wavelet transform to practical data is beyond the scope of this paper.  In a future work we intend to 
examine the localisation properties of spherical wavelets constructed in our framework using numerical simulations and to apply these to analyse the anisotropies of the \cmbtext\ (\cmb).  Spherical wavelets are a powerful tool for analysing the \cmb, from which we are able to learn a great deal about the physics of the early universe \cite{mcewen:2004,mcewen:2006b,mcewen:2006}.

\bibliographystyle{IEEEtran}
\bibliography{cswt2_bibabbrv,cswt2_bib}
%



%

\begin{biography}[{\includegraphics[width=1in,height=1.25in,clip,keepaspectratio]%
{}}]{Jason McEwen}
was born in Wellington, New Zealand, in August 1979.
He received a B.E.\ (Hons) degree in Electrical and Computer Engineering
from the University of Canterbury, New Zealand, in 2002.  

Currently, he is working towards a Ph.D.\ degree at the Astrophysics Group,
Cavendish Laboratory, Cambridge.  His area of interests include
spherical wavelets, the cosmic microwave background and wavelet based reflectance and illumination in computer graphics.
\end{biography}

\begin{biography}[{\includegraphics[width=1in,height=1.25in,clip,keepaspectratio]%
{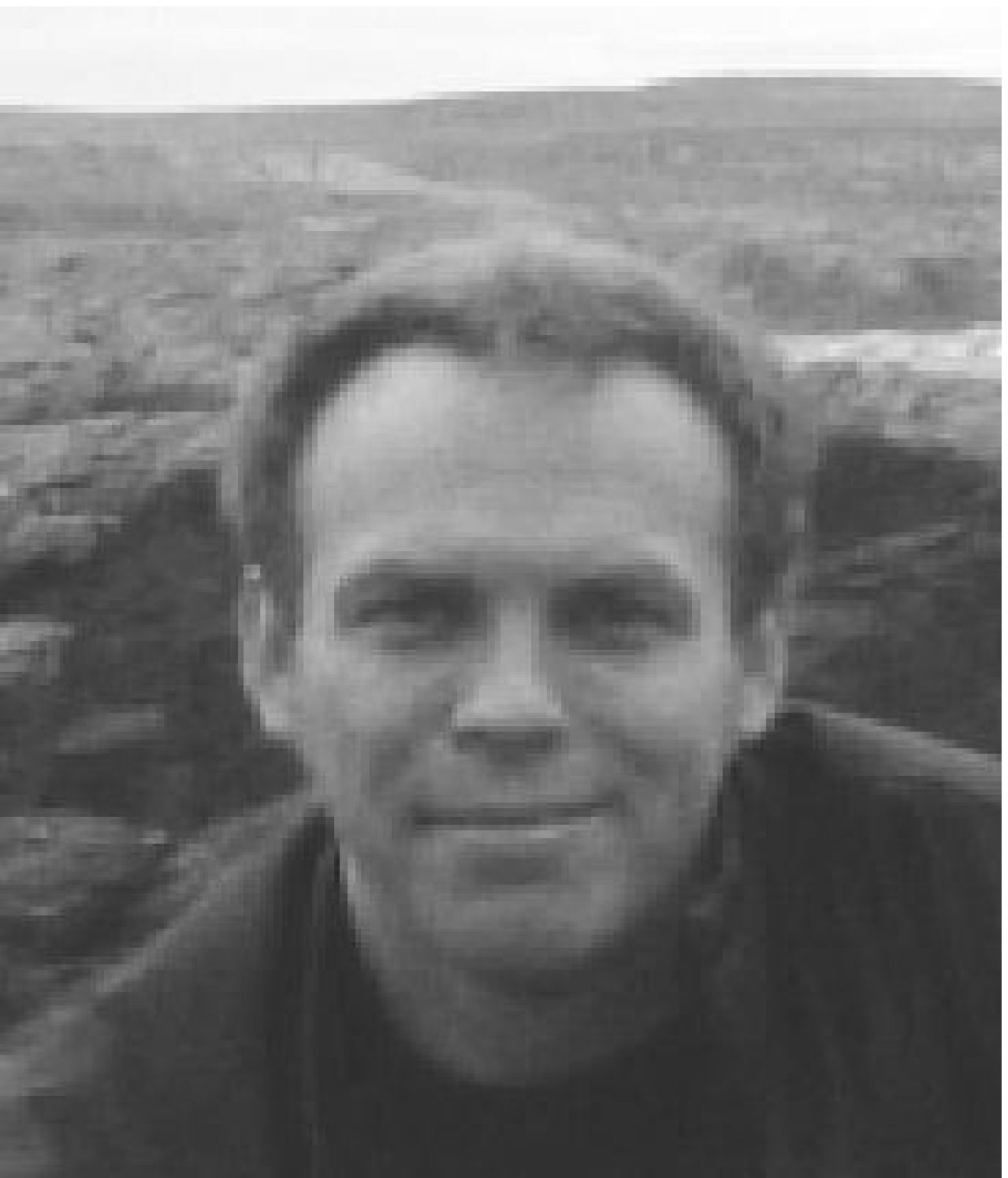}}]{Michael Hobson} 
was born in Birmingham, England, in September 1967. He
received the B.A.\ degree in natural sciences with honours and
the Ph.D.\ degree in astrophysics from the University of
Cambridge, England, in 1989 and 1993 respectively.

Since 1993, he has been a member of the Astrophysics Group of the
Cavendish Laboratory at the University of Cambridge, where he has been
a Reader in Astrophysics and Cosmology since 2003. His research
interests include theoretical and observational cosmology,
particularly anisotropies in the cosmic microwave background,
gravitation, Bayesian analysis techniques and theoretical optics. 
\end{biography}


\begin{biography}[{\includegraphics[width=1in,height=1.25in,clip,keepaspectratio]%
{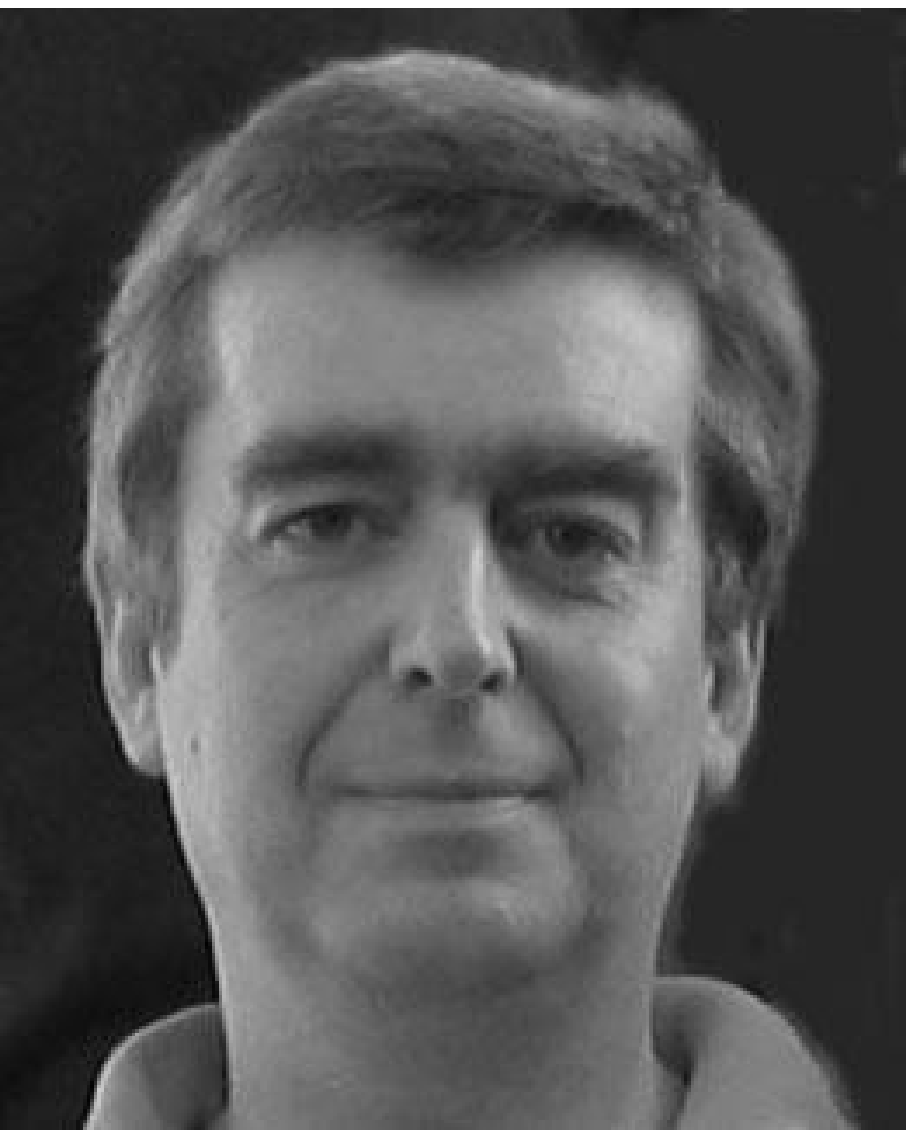}}]{Anthony Lasenby}
was born in Malvern, England, in June 1954. He
received a B.A.\ then M.A.\ from the University of Cambridge in
1975 and 1979, an M.Sc.\ from Queen Mary College, London in 1978
and a Ph.D.\ from the University of Manchester in 1981. 

His Ph.D.\
work was carried out at the Jodrell Bank Radio Observatory
specializing in the Cosmic Microwave Background, which has been a
major subject of his research ever since. After a brief period at
the National Radio Astronomy Observatory in America, he moved
from Manchester to Cambridge in 1984, and has been at the
Cavendish Laboratory Cambridge since then. He is currently Head
of the Astrophysics Group and the Mullard Radio Astronomy
Observatory in the Cavendish Laboratory, and a Deputy Head of the
Laboratory. His other main interests include theoretical physics
and cosmology, the application of new geometric techniques in
computer graphics and electromagnetic modelling, and statistical
techniques in data analysis.
\end{biography}




\end{document}